\documentclass[12pt]{iopart}

\usepackage{iopams}
\usepackage{graphicx}
\usepackage{xcolor}
\begin{document}

\title[]{Measurement of neutron induced reaction cross-section of tantalum with covariance analysis}

\author{Mahima Upadhyay$^1$, Mahesh Choudhary$^1$, Namrata Singh$^1$, Punit Dubey$^1$, Shweta Singh$^1$, Sriya Paul$^1$, Utkarsha Mishra$^1$, G. Mishra$^2$, G. Mohanto$^2$, Sukanya De$^2$, L. S. Danu$^2$, B. Lalremruata$^3$, Ajay Kumar$^2$, R. G. Thomas$^2$, A. Kumar$^1$}

\address{$^{1}$Department of Physics, Banaras Hindu University, Varanasi 221 005, India}
\address{$^{2}$Nuclear Physics Division, Bhabha Atomic Research Centre, Mumbai 400 085, India}
\address{$^{3}$Department of Physics, Mizoram University, Tanhril, Aizawl-796004, India}
\ead{ajaytyagi@bhu.ac.in}
\vspace{10pt}
\begin{indented}
\item[\today]
\end{indented}

\begin{abstract}
The current study presents the cross-section measurement of $^{181}$Ta(n,$\gamma$)$^{182}$Ta reaction at 1.37 $\pm$ 0.13, 2.06 $\pm$ 0.14, 2.56 $\pm$ 0.15, and 3.05 $\pm$ 0.17 MeV neutron energies utilizing offline $\gamma$-ray spectroscopy. The neutrons were generated through the $^{7}$Li(p,n)$^{7}$Be reaction. The $^{115}$In(n,n'$\gamma$)$^{115m}$In reaction served as a monitor reaction. The covariance analysis was used to quantify the uncertainties in the measured cross-sections for the first time for the $^{181}$Ta(n,$\gamma$)$^{182}$Ta reaction. The present study provides detailed information on the propagation of uncertainty in the overall result. The required corrections for low energy background neutron and $\gamma$-ray coincidence summing effect have been made in the present measurement. The output is compared with the pre-existing cross-section data from the EXFOR database, evaluated data libraries and theoretical model predictions.
\end{abstract}

%
%
%
%
%

\section{Introduction}
Neutron capture cross-sections are crucial in various research domains, including the safety assessment and design of reactors, nuclear medicine, nuclear security, nuclear waste management and also in the study of nuclear astrophysics and nuclear structure \cite{01,02,03,04,05}. Several elements are studied by neutron induced reactions to apply this science to our best knowledge.
Tantalum is one such element that has significantly contributed to the field of reactor design as a structural material and in the field of nuclear medicine. Tantalum metals and alloys possess exceptional properties such as high melting point of 2996°C and remarkable thermal strength, making them suitable for applications in high-temperature environments \cite{06}. Tantalum serves as a control material in fast neutron reactors. It has been regarded as a potential material for use in the design of accelerator driven systems and is also crucial for future fusion reactors \cite{07}. The capture cross-section of tantalum is primarily necessary for burnup calculations and quick breeder management \cite{08}. The neutron bombardment on tantalum produces $^{182}$Ta radioisotope, which emits 1.12 MeV $\gamma$ radiation. Due to these, $^{182}$Ta is commonly used as a source of $\gamma$ radiation \cite{09}. This radioisotope has been utilized in the form of a radioactive wire for interstitial implants as a substitute for sealed radium sources \cite{10,Sinclair}. Additionally, it is also used as a radioactive source for ophthalmic applicators and bladder implants \cite{09}. Furthermore, the low energy emissions of $^{182}$Ta can be employed to determine the efficiency curve within the energy range of 100 to 260 keV \cite{IntroTaEff}. Therefore, it is requisite to measure the cross-section of neutron induced reaction on tantalum which could give a reliable nuclear data. There are many previous experimental cross-section data on EXFOR database \cite{11,12}, but none of the accessible data offer a comprehensive covariance analysis.

The current investigation centers on the neutron averaged cross-section for $^{181}$Ta(n,$\gamma$)$^{182}$Ta at neutron energies of 1.37, 2.06, 2.56 and 3.05 MeV. The cross-section of $^{115}$In(n,n'$\gamma$)$^{115m}$In standard monitor reaction obtained from the IRDFF-1.05 data library \cite{13} was utilized to conduct the measurement. In the present study, we have assessed the covariance matrix and uncertainties of the neutron averaged cross-section \cite{14,15}. The current experimental data is compared to the cross-sectional data from the evaluated libraries like TALYS-generated Evaluated Nuclear Data Libraries (TENDL-2019) \cite{16}, Japanese Evaluated Nuclear Data Library (JENDL-5) \cite{17}, ENDF/B-VIII.0 \cite{ENDF}. Theoretical nuclear cross-section calculations were performed using TALYS-1.96 \cite{18}. The data has also been compared with the previous experimental data from the EXFOR database \cite{11,12}. 

The present manuscript is divided into seven distinct sections. Section 2 contains information regarding the experiment, while Section 3 describes the procedure for data analysis. Section 4 provides the theoretical framework, while Section 5 contains the results and discussions. Section 6 concludes the discussion.


\section{DETAILS OF EXPERIMENT}

\begin{table*}[b]
    
\begin{center}
\caption{Details regarding foils used in the experiment.}
\begin{indented}
\vspace{1mm} 

\begin{tabular}{ccccc}
\hline
\hline
Foils & Abundance of Isotope & Density & $<$E$_n$$>$ & Weight of sample \\
&(\%)&(gm/cm$^3$)&(MeV)&(mg)\\
\hline
& & & 1.37  & 248.6 \\
Ta & 99.98 $\pm$ 0.0003 & 16.65 & 2.06  & 218.1 \\
& & & 2.56  & 244.0 \\
& & & 3.05  & 217.4 \\
\hline
& & & 1.37 & 30.6 \\
In & 95.71 $\pm$ 0.0520 & 7.31 & 2.06  & 36.3 \\
& & & 2.56  & 100.0 \\
& & & 3.05  & 46.1 \\

\hline
\hline

\end{tabular}
\end{indented}
\end{center}
\end{table*}

\subsection{Neutron Production}

The experiment was executed at the FOTIA (Folded Tandem Ion Accelerator) Facility, BARC, Mumbai, India. For the present measurement, a lithium target was bombarded by a proton beam of energies 3.3, 4.0, 4.5, 5.0 MeV with an energy spread of 0.02 MeV. This bombardment resulted in the production of neutrons by the reaction $^{7}$Li(p,n)$^{7}$Be, where the Q-value was -1.644 MeV and the threshold energy was 1.880 MeV. The target foil was placed at a distance of 5 mm from the lithium target.

The current experiment involves a proton beam energy surpassing the (p,n$_1$) threshold at 2.37 MeV, which corresponds to the reaction threshold of the first excited state of $^{7}$Be. As a result, it is crucial to deduct low energy neutron contributions from the (p,n$_1$) and breakup neutron contributions from the $^7$Li(p,n+$^3$He)$^4$He reaction. The effect of $^7$Li(p,n+$^3$He)$^4$He is insignificant. Furthermore, due to the continuous nature of the proton beam, it was not possible to utilize the time of flight technique to assess the energy spectrum of the neutron flux. In order to address these two issues, we obtained the neutron flux energy spectrum using the simulation code, EPEN (Energy of Proton Energy of Neutron) \cite{20}. The code is intended to accommodate incident proton energies between 1.88 and 7 MeV. The code generated the individual spectra (p,n$_0$) and (p,n$_1$) for a given incident proton energy. Further details about the EPEN code can be found in Ref. \cite{20,21}.

The energy spectrum of neutron flux for (p,n$_0$) and (p,n$_1$) reactions, simulated by using the EPEN code, is illustrated in Figure 1 for each of the four energies. As mentioned in Ref. \cite{22}, the average energy of the neutron spectrum generated by the $^{7}$Li(p,n$_0$)$^{7}$Be reaction was calculated utilizing the equation provided as follows:

\begin{equation}
<E_n> = \frac{\int_{E_{min}}^{E_{max}} \phi_0(E)EdE}{\int_{E_{min}}^{E_{max}} \phi_0(E)dE} 
\end{equation}

where,\\
$\phi_0$ = neutron flux of (p,n$_0$) spectrum.\\
\textit{E$_{min}$} = 1.003, 1.672, 2.139 and 2.602. \\ 
\textit{E$_{max}$} = 1.714, 2.424, 2.928 and 3.432.\\

The integration limits for the proton energies are denoted as E$_{min}$ and E$_{max}$. The resultant neutron energies with their uncertainties are 1.37 $\pm$ 0.13, 2.06 $\pm$ 0.14 and 2.56 $\pm$ 0.15, 3.05 $\pm$ 0.17 MeV. The uncertainties were calculated by employing the full width at half maximum (FWHM) of the distribution of neutron energy.

\begin{figure}
\begin{center}
\includegraphics[width= 8.4 cm]{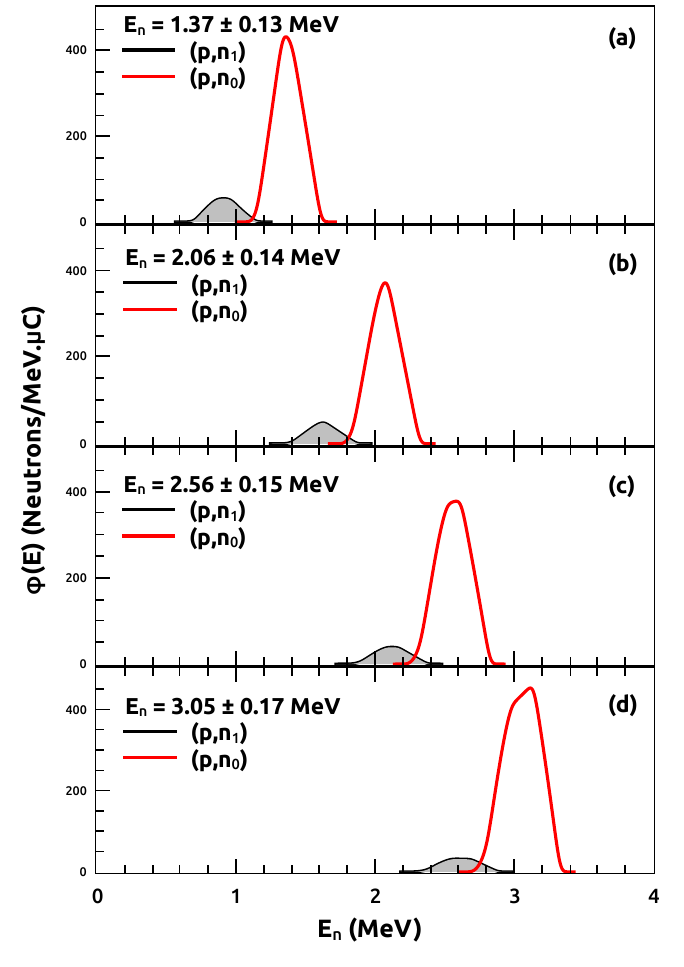}
\caption{Neutron flux for (p,n$_0$) and (p,n$_1$) generated from EPEN code for E$_p$ = 3.3 $\pm$ 0.02, 4.0 $\pm$ 0.02, 4.5 $\pm$ 0.02, 5.0 $\pm$ 0.02 MeV.} 
\end{center}
\end{figure}


\subsection{Offline $\gamma$-ray spectrometry}

 The samples of Ta and In foil were prepared in four separate sets. The weight of the foils was measured using a micro-balance device with an accuracy of 0.1 mg. The details of the weight of the sample are briefly provided in Table 1 for each corresponding energy. The Ta and In foils were each encased in 14 $\mu$m thick aluminium foil to suppress the radioactive cross-contamination among the foils and their surroundings. The samples were exposed to radiation by positioning the stack of "In-Ta" with dimensions of 10 x 10 mm$^2$ at a zero-degree angle relative to the beam direction.

The samples were cooled and transferred to the counting chamber subsequent to irradiation. The details regarding irradiation, cooling, and counting time are provided in Table 2. We have utilized p-type high-purity germanium (HPGe) detector of relative efficiency 30\% for offline $\gamma$-ray spectroscopy to measure the induced activity. The data acquisition was performed using the CAMAC based LAMPS program. The decay data of $\gamma$-ray energy and associated uncertainties were obtained from the online database, NuDat 3.0 \cite{23}, and are presented in Table 3.

\section{Data Analysis} 

\begin{table*}[b]
    
\begin{center}
\caption{Details of irradiation, cooling and counting time.}
\begin{indented}
\vspace{1mm} 

\begin{tabular}{ccccc}
\hline
\hline
Reaction & $<$E$_n$$>$ & Irradiation time  & Cooling time & Counting time \\
&(MeV)&(s)& (s) & (s)\\
\hline
& 1.37 $\pm$ 0.13 & 36240 & 75060 & 18757 \\
$^{181}$Ta(n,$\gamma$)$^{182}$Ta ~ & 2.06 $\pm$ 0.14 & 41460 & 24000 & 12814 \\
& 2.56 $\pm$ 0.15 & 40800 & 7800 & 34131 \\
& 3.05 $\pm$ 0.17 & 26040 & 30300 & 32260 \\
\hline

& 1.37 $\pm$ 0.13 & 36240 & 31860 & 1676 \\
$^{115}$In(n,n'$\gamma$)$^{115m}$In ~ & 2.06 $\pm$ 0.14 & 41460 & 2040 & 711 \\
& 2.56 $\pm$ 0.15 & 40800 & 3780 & 1558 \\
& 3.05 $\pm$ 0.17 & 26040 & 18420 & 1837 \\
\hline
\hline
\end{tabular}
\end{indented}
\end{center}
\end{table*}


\begin{table*}
    
\begin{center}
\caption{Decay data of the interested nuclides in the present work.}
\begin{indented}
\vspace{1mm} 

\begin{tabular}{ccccc}
\hline
\hline
Reaction & Product nuclide & Half life & E$_\gamma$ & I$_\gamma$ \\
&  & (h) & (keV) & (\%)  \\
\hline
$^{181}$Ta(n,$\gamma$)$^{182}$Ta & $^{182}$Ta & 2753.76 $\pm$ 2.880 ~ & 1121.290 $\pm$ 0.003 ~ & 35.24 $\pm$ 0.08 \\
$^{115}$In(n,n'$\gamma$)$^{115m}$In & $^{115m}$In & 4.486 $\pm$ 0.004 ~ & 336.241 $\pm$ 0.025 ~ & 45.90 $\pm$ 0.10 \\

\hline
\hline

\end{tabular}
\end{indented}
\end{center}
\end{table*}

\begin{table*}
\begin{center}
\caption{Efficiency of the detector w.r.t point source and sample shape geometry and correction factor for coincidence summing effect.}
\vspace{2mm}
\begin{indented}
\begin{tabular}{cccccc}
\hline
\hline

$E_{\gamma}$ (keV)&$I_{\gamma}$&Counts (C)&K$_{s}$&$\varepsilon_{p}$ &$\varepsilon$ \\
\hline
\vspace{1mm}
121.78 & 0.2853 $\pm$ 0.0016 & 239969 & 1.1765 ~ & 0.05758 ~ & 0.04356 $\pm$ 0.00093 \\
\vspace{1mm}
244.69 & 0.0755 $\pm$ 0.0004 & 44797 & 1.2450 ~ & 0.04298 ~ & 0.03198 $\pm$ 0.00088 \\
\vspace{1mm}
443.96 & 0.0282 $\pm$ 0.0001 & 10690 & 1.2200 ~ & 0.02684 ~ & 0.02010 $\pm$ 0.00057 \\
\vspace{1mm}
778.90 & 0.1293 $\pm$ 0.0008 & 27794 & 1.1768 ~ & 0.01472 ~ & 0.01111 $\pm$ 0.00021 \\
\vspace{1mm}
964.05 & 0.1451 $\pm$ 0.0007 & 28149 & 1.1043 ~ & 0.01246 ~ & 0.00944 $\pm$ 0.00015 \\
\vspace{1mm}
1112.07 & 0.1367 $\pm$ 0.0008 & 23940  & 1.0474 ~ & 0.01067 ~ & 0.00809 $\pm$ 0.00015 \\
\vspace{1mm}
1408.01 & 0.2087 $\pm$ 0.0009 & 29578  & 1.0725 ~ & 0.00884 ~ & 0.00673 $\pm$ 0.00011 \\

\hline
\hline
\end{tabular}
\end{indented}
\end{center}
\end{table*}

\begin{figure}
\begin{center}
\includegraphics[width=8.0 cm]{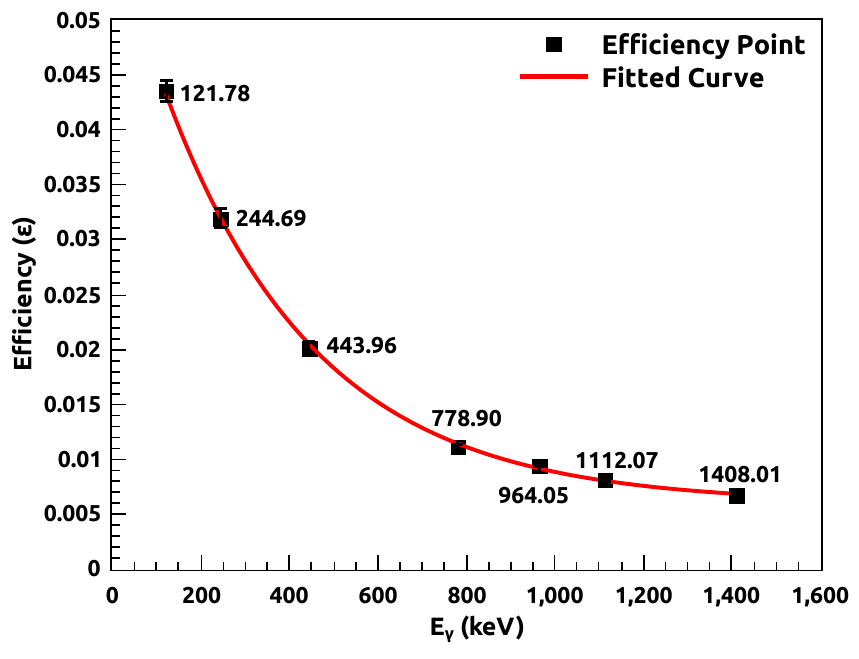}
\caption{Interpolated efficiencies of the detector.} 
\end{center}
\end{figure}


\begin{table*}
    
\begin{center}
\caption{Efficiency of the specific $\gamma$-ray energy of the product radionuclides of the sample and monitor with their uncertainties and correlation matrix.}
\begin{indented}
\vspace{2mm} 

\begin{tabular}{ccccc}
\hline
\hline
Reaction & E$_\gamma$ & Efficiency & Correlation Matrix  \\

\hline
$^{181}$Ta(n,$\gamma$)$^{182}$Ta  ~ & 1121.290 $\pm$ 0.003 ~ & 0.00797 $\pm$ 0.00010  & 1 & \\
$^{115}$In(n,n'$\gamma$)$^{115m}$In  ~ & 336.241 $\pm$ 0.025 ~ & 0.02587 $\pm$ 0.00044 & 0.043 & 1\\

\hline
\hline

\end{tabular}
\end{indented}
\end{center}
\end{table*}

\subsection{Measurement of $\gamma$-ray activity}

The efficiency of the p-type high-purity germanium detector at different characteristic $\gamma$-ray energies was calibrated utilizing a standard $^{152}$Eu point source with a half-life of 13.517 $\pm$ 0.009 years and initial activity of 6614.71 $\pm$ 81.33 Bq as of October 1, 1999. The formula assessed to determine the detector efficiency for the point source is briefly described in Ref. \cite{24,25}. The correction for the coincidence summing effect is incorporated in the calibration of efficiency. We have computed it using a Monte Carlo simulation program called EFFTRAN \cite{26}. Additionally, the efficiency with respect to the point source geometry was transferred to sample shape geometry using the EFFTRAN code. In our previous publication \cite{MahimaMo}, we presented a comprehensive elucidation of the coincidence summing effect and its correction in the measurement. Table 4 provides the efficiency of the detector with respect to the point source ($\varepsilon_{p}$), sample shape geometry ($\varepsilon$) and correction factor for the coincidence summing effect (K$_{s}$).

The efficiency of the detector for the specific energy of the $\gamma$-ray emitted by the product radionuclides was ascertained by interpolating the detector efficiency. The calibration curve of the detector efficiency is shown in Figure 2. We have used the formula provided in Ref. \cite{27} to estimate the uncertainties of the efficiencies for the characteristic $\gamma$-ray energy where we have propagated the covariances of the three fitting parameters to the efficiencies of the product radionuclides. The efficiency of the specific $\gamma$-ray energy of the of the product radionuclides of the sample and monitor with their uncertainties and correlation matrix are given in Table 5.



\begin{table}
\begin{center}

\caption{Group-wise quantities for neutron flux spectrum, uncertainty and correlation coefficients of monitor reaction between different energies.}
\footnotesize

\begin{tiny}
    \rotatebox{90}{

\begin{tabular}{cccccccccccccccccccccc}

\br

$l$& $h$ &$E_{h,min}$ & $E_{h,max}$ & $\phi_{l,h}$ / $\sum\phi_{l,h}$ & $\Delta\sigma_h$ &\multicolumn{16}{c}{Correlation coefficient Cor ($\sigma_h,\sigma_i$)}\\
& & (MeV) & (MeV) & & (\%) & 1.0 & 1.2 & 1.4 & 1.6 & 1.6 & 1.8 & 2.0 & 2.2 & 2.2 & 2.4 & 2.6 & 2.8 & 2.6 & 2.8 & 3.0 & 3.2 \\
\mr

1& 1 & 1.0 & 1.2 & 0.0387 & 3.06 & 1.00 &&&&&&&&&&&&&&&\\
&2 & 1.2 & 1.4 & 0.5663 & 3.02 & 0.81 & 1.00 &&&&&&&&&&&&&&\\
&3& 1.4 & 1.6 & 0.3893 & 2.93 & 0.48 & 0.82 & 1.00 &&&&&&&&&&&&&\\
&4 & 1.6 & 1.8 & 0.0043 & 2.81 & 0.21 & 0.46 & 0.84 & 1.00 &&&&&&&&&&&&\\
2& 5 & 1.6 & 1.8 & 0.0016 & 2.81 & 0.21 & 0.46 & 0.84 & 1.00 & 1.00 &&&&&&&&&&&\\
&6 & 1.8 & 2.0 & 0.2684 & 2.67 & 0.15 & 0.21 & 0.54 & 0.86 & 0.86 & 1.00 &&&&&&&&&&\\
&7 & 2.0 & 2.2 & 0.6087 & 2.63 & 0.25 & 0.17 & 0.26 & 0.51 & 0.51 & 0.83 & 1.00 &&&&&&&&&\\
&8 & 2.2 & 2.4 & 0.1211 & 2.62 & 0.34 & 0.27 & 0.17 & 0.21 & 0.21 & 0.49 & 0.85 & 1.00 &&&&&&&&\\
3&9 & 2.2 & 2.4 & 0.0845 & 2.62 & 0.34 & 0.27 & 0.17 & 0.21 & 0.21 & 0.49 & 0.85 & 1.00 & 1.00 &&&&&&&\\
&10 & 2.4 & 2.6 & 0.5254 & 2.59 & 0.33 & 0.34 & 0.25 & 0.17 & 0.17 & 0.27 & 0.56 & 0.87 & 0.87 & 1.00 &&&&&&\\
&11 & 2.6 & 2.8 & 0.3808 & 2.57 & 0.26 & 0.33 & 0.36 & 0.29 & 0.29 & 0.23 & 0.31 & 0.57 & 0.57 & 0.87 & 1.00 &&&&&\\
&12 & 2.8 & 3.0 & 0.0092 & 2.45 & 0.22 & 0.27 & 0.40 & 0.41 & 0.41 & 0.31 & 0.22 & 0.32 & 0.32 & 0.61 & 0.90 & 1.00 &&&&\\
4&13 & 2.6 & 2.8 & 0.0150 & 2.57 & 0.26 & 0.33 & 0.36 & 0.29 & 0.29 & 0.23 & 0.31 & 0.57 & 0.57 & 0.87 & 1.00 & 0.90 & 1.00 &&&\\
&14 & 2.8 & 3.0 & 0.3235 & 2.45 & 0.22 & 0.27 & 0.40 & 0.41 & 0.41 & 0.31 & 0.22 & 0.32 & 0.32 & 0.61 & 0.90 & 1.00 & 0.90 & 1.00 &&\\
&15 & 3.0 & 3.2 & 0.5229 & 2.37 & 0.28 & 0.26 & 0.34 & 0.42 & 0.42 & 0.43 & 0.34 & 0.26 & 0.26 & 0.33 & 0.54 & 0.78 & 0.54 & 0.78 & 1.00 &\\
&16 & 3.2 & 3.4 & 0.1384 & 2.38 & 0.33 & 0.30 & 0.26 & 0.27 & 0.27 & 0.36 & 0.43 & 0.39 & 0.39 & 0.28 & 0.23 & 0.33 & 0.23 & 0.33 & 0.73 & 1.00\\

\br

\end{tabular}}
\end{tiny}
\end{center}
\end{table}


\subsection{Cross-section for neutron spectrum averaged monitor reaction }

In this study, $^{115}$In(n,n'$\gamma$)$^{115m}$In reaction was employed as a reference monitor reaction. The cross-section and covariance data for this reaction were sourced from IRDFF-1.05 library. 
The point-wise monitor cross-section ($\sigma_m$(E)) is enfolded by the (p,n$_0$) neutron flux energy spectrum obtained by the EPEN \cite{20} code using the following equation,

\begin{equation}
<\sigma_m> = \frac{\int_{E_{min}}^{E_{max}} \phi_0(E)\sigma_m(E)dE}{\int_{E_{min}}^{E_{max}} \phi_0(E)dE} 
\end{equation}

The group-wise neutron flux spectrum was calculated by,

\begin{equation}
\phi_{l,h} = {\int_{E_{h,min}}^{E_{h,max}} \phi_l(E)dE}
\end{equation}

which satisfies $\Sigma$$\phi_{l,h}$ = 1,  where \textit{l} = 1,  2, 3 and 4 are  classified for $<$$E_n$$>$ = 1.37, 2.06, 2.56, and 3.05 MeV respectively. The IRDFF-1.05 library contains energy group boundaries, which consist of \textit{h} energy groups for each neutron energy \textit{l}. \textit{E$_{h,min}$} and \textit{E$_{h,max}$} denote the lower and upper boundaries, respectively, for each \textit{h$^{th}$} energy group. \textit{h} groups are specified as 1–4, 5–8, 9–12, and 13-16 for  $<$$E_n$$>$ = 1.37, 2.06, 2.56, and 3.05 MeV respectively. The group-wise quantities for neutron flux spectrum, uncertainty and correlation coefficients of monitor cross-sections is listed in Table 6. The equation used to propagate the covariance matrix obtained from the IRDFF-1.05 library to the averaged monitor cross-section is as follows:

\begin{equation}
Cov(<\sigma_m>_l,<\sigma_m>_j)= \sum_{h=1}^N\sum_{l=1}^N\phi_{l,h}Cov(\sigma_h,\sigma_i)\phi_{j,i}
\end{equation}

We calculated the correlation coefficients from the propagated covariance matrix of the averaged monitor cross-section. The monitor cross-section for the average neutron energy, their uncertainty and correlation coefficients are mentioned in Table 7.

\begin{table*}
\begin{center}
\caption{Monitor cross-section for the average neutron energy, their uncertainty and correlation coefficients.}
\vspace{2mm}
\begin{indented}
\begin{tabular}{ccccccc}
\br
$<$E$_n$$>$ & $<$$\sigma_m$$>$ & $\bigtriangleup$$<$$\sigma_m$$>$&\multicolumn{4}{c}{Correlation matrix}\\
 (MeV)&($\sigma$ $\pm$ $\bigtriangleup\sigma$)&&\\
\mr
\vspace{1mm}
1.37 $\pm$ 0.13 & ~ 143.46  $\pm$ 4.33 & ~ 3.02 & ~ 1.000 \\
\vspace{1mm}
2.06 $\pm$ 0.14 & ~ 287.42 $\pm$ 7.55 & ~ 2.63 & ~ 0.252 & ~ 1.000  &\\
\vspace{1mm}
2.56 $\pm$ 0.15 & ~ 342.49 $\pm$ 8.87 & ~ 2.59 & ~ 0.313 & ~ 0.470 & ~ 1.000\\
\vspace{1mm}
3.05 $\pm$ 0.17 & ~ 339.92 $\pm$ 8.05 & ~ 2.37 & ~ 0.305 & ~ 0.341 & ~ 0.504 & ~ 1.000\\

\br
\end{tabular}
\end{indented}
\end{center}
\end{table*}

\subsection{Correction for low energy background neutrons}

The present proton energies exceed the reaction threshold for the first excited state of $^{7}$Be. The neutrons produced from the reaction $^{7}$Li(p,n$_0$)$^{7}$Be  were contaminated by a secondary group of lower energy neutrons produced from the reaction $^{7}$Li(p,n$_1$)$^{7}$Be. This low-energy background neutron contribution must be subtracted in order to measure the neutron-induced reaction cross-section precisely. This has been taken into account and computed utilizing following equation,

\begin{equation}
L_{cor} = 1 - \frac{\int_{E_{min}}^{E_{max}} \phi_1(E)(\sigma_x(E))dE}{\int_{E_{min}}^{E_{max}} \phi(E)(\sigma_x(E))dE} 
\end{equation}

where, 

$\phi$ represents the total neutron flux computed by the EPEN code. The neutron flux spectrum of the (p,n$_1$) and (p,n$_0$) are denoted by $\phi_1$ and $\phi_0$ ($\phi$ =  $\phi_1$ + $\phi_0$). For neutron spectra (p,n$_1$), the integrating limits are $E_{min}$ = 0.565, 1.248, 1.719, 2.184 and $E_{max}$ = 1.252, 1.972, 2.481, 2.987. For neutron spectra (p,n$_0$), the integrating limits are $E_{min}$ = 1.003, 1.672, 2.139, 2.602 and $E_{max}$ = 1.714, 2.424, 2.928, 3.432 for proton energies 3.3, 4.0, 4.5 and 5.0 MeV respectively. 
$\sigma_x$ is the cross-section of $^{181}$Ta(n,$\gamma$)$^{182}$Ta and $^{115}$In(n,n'$\gamma$)$^{115m}$In reactions taken from the IRDFF-1.05 library \cite{13}. The correction factor used in the present measurement for low energy background neutrons is listed in Table 8.

\begin{table*}
    
\begin{center}
\caption{Correction factor calculated in the present measurement for low energy background neutrons. }
\begin{indented}
\vspace{1mm} 

\begin{tabular}{ccc}
\br
Sample & $<$E$_n$$>$ & $L_{cor}$  \\
&(MeV)& \\
\mr
& 1.37 $\pm$ 0.13 ~ & 0.873 ~   \\
Ta & 2.06 $\pm$ 0.14 ~ & 0.854 ~  \\
& 2.56 $\pm$ 0.15 ~ & 0.883 ~  \\
& 3.05 $\pm$ 0.17 ~ & 0.912 ~  \\
\hline

& 1.37 $\pm$ 0.13 ~ & 0.957 ~  \\
In  & 2.06 $\pm$ 0.14 ~ & 0.923 ~  \\
& 2.56 $\pm$ 0.15 ~ & 0.923 ~   \\
& 3.05 $\pm$ 0.17 ~ & 0.936 ~ \\
\br
\end{tabular}
\end{indented}
\end{center}
\end{table*}


\subsection{Quantification of cross-section and its uncertainty}

In the present experiment, the spectrum averaged cross-section ($\sigma_t$) for $^{181}$Ta(n,$\gamma$)$^{182}$Ta was quantified by using the following equation, 

\begin{equation}
<\sigma_t> = <\sigma_{m}>\times\eta
\times\frac{A_{t}N_{m}I_{\gamma(m)}f_{m}}{A_{m}N_{t}I_{\gamma(t)}f_{t}} \times\frac{L_{cor(t)}}{L_{cor(m)}}
\end{equation}

where,\\

$\eta$ is ratio of monitor efficiency to sample efficiency, $\sigma_{m}$ is cross-section of neutron spectrum averaged reference monitor reaction, $A_t$ and $A_m$ are total number of counts from the radioisotope in target and monitor foil, $N_m$ and $N_t$ are total number of atoms in the monitor and target foil, $I_{\gamma(m)}$ and $I_{\gamma(t)}$ are intensity of the specific $\gamma$–ray energy of the radioisotope in monitor and target foil, $f_m$ and $f_t$ are timing factor for the monitor and target foil, $L_{cor(t)}$ and $L_{cor(m)}$ are correction factor for low energy background neutrons for target and monitor foil.

The timing factor was calculated by the following formula,
\begin{equation}
f = (1-e^{-{\lambda}t_{i}}){e^{-{\lambda}t_{u}}}(1-e^{-{\lambda}t_{w}})/{\lambda}
\end{equation}

where, $t_{i}$ is irradiation time, $t_{u}$ is counting time, $t_{w}$ is cooling time, ${\lambda}$ is decay constant\\

The uncertainties in the measured cross-sections were determined by incorporating the fractional uncertainties of the several parameters like $\gamma$-ray photo-peak counts, $\gamma$-ray intensity, efficiency ratio, reference reaction cross-section, timing factor and number of atoms. Fractional uncertainties in several attributes associated in the present work for each energy is mentioned in Table 9. We have calculated the fractional variance and covariance according to the method described in section 4.1.4 of Ref. \cite{28}. The cross-sections of $^{181}$Ta(n,$\gamma$)$^{182}$Ta reaction with their corresponding uncertainties and correlation matrix are presented in Table 10.


\begin{table*}[b]
    
\begin{center}
\caption{Fractional uncertainties (\%) in several attributes associated in the present work for each energy. }
\begin{indented}

\begin{tabular}{ccccc}
\br
Parameters &$<$E$_n$$>$= & $<$E$_n$$>$= & $<$E$_n$$>$= & $<$E$_n$$>$= \\
 & 1.37 MeV& 2.06 MeV& 2.56 MeV & 3.05 MeV \\

\mr
\textit{A$_{t}$} &6.967& 7.738&  7.955 &  9.325 \\
\textit{A$_{m}$} &2.917& 1.103&  0.623 &  1.007 \\
\textit{N$_{t}$} &0.040& 0.045& 0.040  & 0.046 \\
\textit{N$_{m}$} &0.331& 0.280& 0.113 & 0.223 \\
\textit{f$_{t}$} &0.103& 0.104& 0.104 & 0.104 \\
\textit{f$_{m}$} &0.087& 0.023& 0.015 & 0.025 \\
\textit{I$_{t}$} &0.227& 0.227&  0.227 &  0.227 \\
\textit{I$_{m}$} &0.217& 0.217& 0.217 & 0.217 \\
\textit{$\eta$} &2.100& 2.100& 2.100 & 2.100 \\
\textit{$\sigma_{m}$} &3.018& 2.626&  2.589 &  2.368 \\
Total Error (\%) &8.41& 8.52& 8.65 & 9.90\\

\br

\end{tabular}
\end{indented}
\end{center}
\end{table*}


\subsection{Covariance analysis} 


\begin{table*}
\begin{center}
\caption{Cross-sections of $^{181}$Ta(n,$\gamma$)$^{182}$Ta reaction with their corresponding uncertainties and correlation matrix.} 
\vspace{2mm}
\begin{indented}
\begin{tabular}{cccccc}
\br
$<$E$_n$$>$ &Present data (mb)&\multicolumn{4}{c}{Correlation matrix}\\
 (MeV)&($\sigma$ $\pm$ $\bigtriangleup\sigma$)&&\\
\mr
\vspace{1mm}
1.37 $\pm$ 0.13 & ~ 123.94 $\pm$ 10.42 & ~ 1.000 \\
\vspace{1mm}
2.06 $\pm$ 0.14 & ~ 82.62 $\pm$ 7.04 & ~ 0.091 & ~ 1.000  &\\
\vspace{1mm}
2.56 $\pm$ 0.15 & ~ 57.23 $\pm$ 4.95 & ~ 0.095 & ~ 0.104 & ~ 1.000\\
\vspace{1mm}
3.05 $\pm$ 0.17 & ~ 48.12 $\pm$ 4.76 & ~ 0.080 & ~ 0.078 & ~ 0.088 & ~ 1.000\\

\br
\end{tabular}
\end{indented}
\end{center}
\end{table*}

Covariance analysis is a robust statistical approach used to evaluate the uncertainty and correlations between different experimental observables. Covariance analysis can provide insight into the relationships between different parameters evaluated in experiments, helping to comprehend their interdependencies. The fractional variance and covariance matrix are formed by the method mentioned in section 4.1.4 of Ref. \cite{28}. The correlation coefficient ($a_x$,$a_y$) is the correlation between the two particular source of uncertainties denoted as $a_x$ and $a_y$. The correlation coefficients must fall within the range of -1 to +1 \cite{27,28}. The uncorrelated coefficients are displayed as Cor($a_x$,$a_y$) = 0 and additionally, the fully correlated coefficient Cor($a_x$,$a_y$) = 1 signifies a perfect correlation between $a_x$ and $a_y$. 

\begin{equation}
Cor(a_x,a_y)= \frac{Cov(a_x,a_y)}{\Delta a_x \Delta a_y}
\end{equation}

The correlation between two parameters is determined with the help of covariance matrix utilizing the aforementioned equation. In the present measurement, we have investigated the cross-section of the $^{181}$Ta(n,$\gamma$)$^{182}$Ta  reaction at 1.37 $\pm$ 0.13, 2.06 $\pm$ 0.14 and 2.56 $\pm$ 0.15, 3.05 $\pm$ 0.17 MeV neutron energies. A correlation exists between the cross-section of our reaction at all four neutron energies and efficiency of the detector as we have employed the same HPGe detector for counting of all four irradiated samples. In addition, we have also computed the correlation between the averaged monitor cross-section for each of the four specific neutron energies.

\section{Details of the Theoretical Evaluation}

The statistical nuclear model code TALYS-1.96 \cite{18} has been used to perform theoretical calculations for the reaction $^{181}$Ta(n,$\gamma$)$^{182}$Ta in the energy range 1-4 MeV. This code can compute the reaction cross-section as a function of input energies up to 200 MeV, considering the impacts of level density parameters and various reaction mechanisms, such as compound nucleus, pre-equilibrium emission, and direct reaction. The TALYS-1.96 code provides all potential outward reaction channels for a particular projectile and target system. In this work, six distinct level density models (LD models) included in the code were used to simulate the nuclear reaction cross-sections. They are listed as the constant temperature Fermi gas model (LDM1) \cite{33}, back-shifted Fermi gas model (LDM2) \cite{34}, generalized
superfluid model (LDM3) \cite{35,36}, skyrme HFB (LDM4), gogny HFB (LDM5) \cite{37}, temperature dependent gogny HFB (LDM6) \cite{38}.

\begin{figure}[b]
\begin{center}
\includegraphics[width=8 cm]{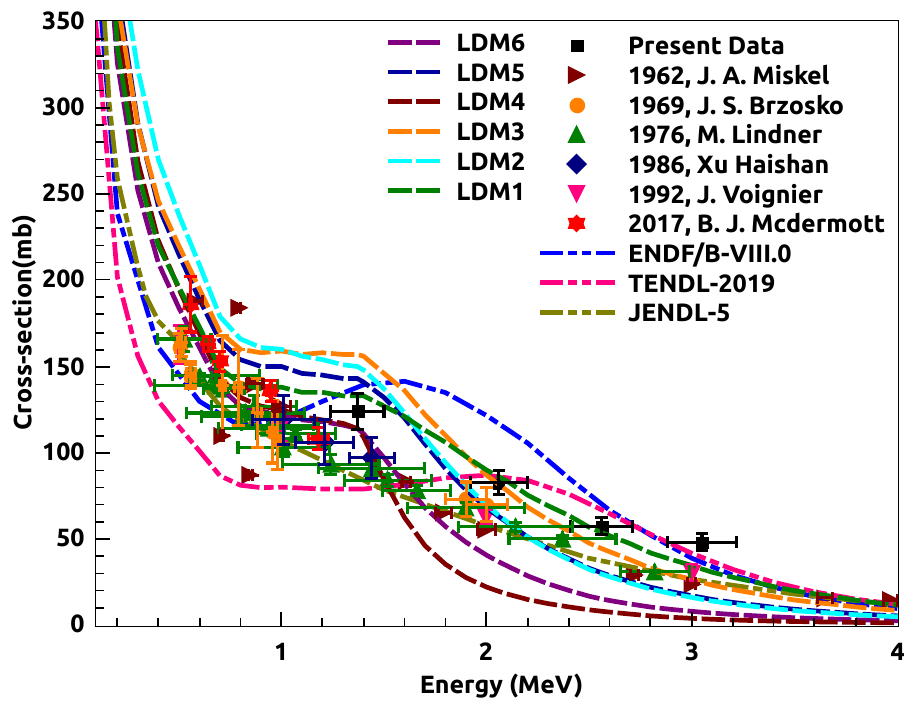}
\caption{Cross-section of $^{181}$Ta(n,$\gamma$)$^{182}$Ta reaction studied in the present work compared with different level density models, EXFOR database and different evaluated data libraries.} 
\end{center}
\end{figure}

\section{Results and Discussions}

The cross-sections of the $^{181}$Ta(n,$\gamma$)$^{182}$Ta reaction, together with their uncertainties and correlation matrix, have been determined at neutron energies of 1.37 $\pm$ 0.13, 2.06 $\pm$ 0.14, 2.56 $\pm$ 0.15, and 3.05 $\pm$ 0.17 MeV enumerated in Table 10. In Figure 3, we have compared our experimental cross-sections with the assessed data obtained from EXFOR, JENDL-5 \cite{17}, TENDL-2019 \cite{16}, ENDF/B-VIII.0 \cite{ENDF} and theoretical findings gained from TALYS-1.96 \cite{18}. The overall uncertainty in the measured cross-sections is within the range of 8 to 10\% which is less, compared to the relative uncertainty evaluated by IRDFF-1.05 library. The measured cross-section is shown by black rectangles in Figure 3. 

We have rationalized the results with different evaluated data libraries like JENDL-5 (dark yellow) , TENDL-2019 (pink), ENDF/B-VIII.0 (blue) which are represented in dash-dot-dot-dash lines. The present experimental result is best in agreement with the evaluated data library ENDF/B-VIII.0 except at 2.06 MeV. The present data at 2.06, 2.56 MeV and 3.05 MeV is consistent with the  TENDL-2019 data library excluding the cross-section data point at 1.37 MeV. The different nuclear LD models are indicated as LDM1 (olive), LDM2 (cyan), LDM3 (orange), LDM4 (wine), LDM5 (royal), LDM6 (purple) in dash-dash lines. The present quantified cross-section is in good agreement with the LDM1. All other theoretical results underestimate the current cross-section at 2.56 and 3.05 MeV. The measured cross-section data at 1.37 MeV is close to the data simulated by LDM1, LDM4 and LDM6. 

The measured cross-section follows the trend with the previous experimental results. The data reported by Xu Haishan \textit{et al}. \cite{Haishan86}, J. Voignier \textit{et al}. \cite{Voignier}, B. J. Mcdermott \textit{et al}. \cite{Mcdermott} have used time of flight technique for the evaluation of cross-section whereas the data studied by J. A. Miskel \textit{et al}. \cite{Miskel}, J. S. Brzosko \textit{et al}. \cite{Brzosko}, M. Lindner \textit{et al}. \cite{Lindner} have employed activation technique. The measured cross-section at 2.06 MeV is within the upper limit of the data reported by J. S. Brzosko \textit{et al.} \cite{Brzosko}. The measured cross-sections follow the trend as predicted by theoretical model LDM1 whereas none of the prior measurements follow the trend of any theoretical model. The present experimental data at 1.37 MeV is well supported by the theoretical data (LD models) within range of 0.8 to 1.4 MeV. Additionally, evaluated data libraries i.e. ENDF/B-VIII.0 and TENDL-2019 rises up in the same region which is very intriguing. All pre-existing experimental results shows deviation and are not in agreement with the LD models in this region.

\section{Conclusions} 

In this study, we have measured the cross-section of $^{181}$Ta(n,$\gamma$)$^{182}$Ta reaction in the energy range 1.37 to 3.05 MeV using the neutron activation method and offline $\gamma$-ray spectroscopy. The covariance analysis was used to quantify uncertainties in the measured cross-sections for the first time for this reaction. The uncertainties in the present results are in range of 8-10\% which is less, compared to the previous experimental results of J. S. Brzosko \textit{et al}, Xu Haishan \textit{et al}, J. Voignier \textit{et al}. We have compared the present measurement with the TALYS predictions, evaluated data and available experimental data. It is found that the present data is in trend with the evaluated data of ENDF/B-VIII.0 and theoretical outcome of constant temperature Fermi gas model (LDM1). All prior experimental results have large inconsistencies with the present measured cross-sections.\\


\ack 
The authors express their gratitude to the FOTIA facility team for their exceptional management of the accelerator and assistance throughout the experiment. We would also like to thank TIFR-Target Laboratory for their aid in target preparation. The author, Mahima Upadhyay, would like to acknowledge the UGC Non-NET Fellowship (Sanction No. R/Dev/Sch/UGC Non-NETFellow/2022-23/57524) for providing financial assistance. One of the authors (A. Kumar) is thankful to the IUAC-UGC, Government of India (Sanction No. IUAC/XIII.7/UFR-71353).\\

\end{document}